# Frequency Analysis of Decoupling Capacitors for Three Voltage Supplies in SoC

Mohd Abubakr,
Electronics and Communication Engineering
Gokaraju Rangaraju Institute of Engineering and Technology
Miyapur, Hyderabad, 500072 INDIA
Email id: mohdabubakr@gmail.com

*Abstract*—Reduction in power consumption has become a major criterion of design in modern ICs. One such scheme to reduce power consumption by an IC is the use of multiple power supplies for critical and non-critical paths. To maintain the impedance of a power distribution system below a specified level, multiple decoupling capacitors are placed at different levels of power grid hierarchy. This paper describes about three-voltage supply power distribution systems. The noise at one power supply can propagate to the other power supply, causing power and signal integrity problems in the overall system. Effects such as anti-resonance and remedies for these effects are studied.

Impedance of the three-voltage supply power distribution system is calculated in terms of RLC-model of decoupling capacitors. Further the obtained impedance depends on the frequency; hence brief frequency analysis of impedance is done.

## I. INTRODUCTION

Power has become one of the most important paradigms of design convergence in 45nm and not the performance [2]. This unconventional change has occurred due to the increasing demand for portable applications. A portable application makes use of battery-powered systems and thus appends a constraint of battery life before the designers. Although there has been substantial development in the increase of battery life, the stringent increase doesn't seem to be eminent [7]. Another reason for power becoming a concerning issue is the dissipation levels in high-performance computing using complex architectures. Power consumption in Intel processors has increased exponentially with the generations of process technologies [2]. Recent results show that a 1kg NiCd battery can support a Pentium 4 for less than an 1 hour whereas it can sustain a Centrino notebook for >3 hours. This reflects that reducing power dissipation can increment the overall performance of the system.

In the field of embedded systems, cropping of power consumption has always been a critical design issue. The easiest and effective way to accomplish reduction in power consumption by a particular circuit is to reduce its voltage supply level. Nevertheless any reduction in the supply voltage gives rise to propagation delays in the circuit. One of the techniques to compensate the propagation delays in circuits is shortening the critical paths in the data-path using behavioral transformations such as parallelization and pipelining. However the resulting circuit consumes lower average power while meeting the global throughput constraint at the cost of increased overhead circuit area.

In recent times, the use of multiple on-chip supply voltages has become an attractive technique to reduce the power consumption without causing any delays. The trick here is to allow the modules present at critical paths to use the highest voltage level and the modules residing on non-critical paths to use lower voltages. Delivering the modules on the critical paths with higher voltage satisfies the target timing constraints and supplying lower voltages to the modules present at non-critical paths reduces the energy consumption of the circuit. This scheme tends to result in smaller area overhead compared with the parallel architectures. In this whole process the system frequency is not affected. Practically a system with multiple voltage supplies faces certain problems such as multiple voltage scheduling [6]. Though having multiple voltage supplies on a chip appears attractive, there is a trade-off between cost and area. Hence the availability of two to three voltage supplies is realistic.

This paper is organized as follows. Section II describes about general overview about power distribution networks. Prior work on two-voltage supplies power distribution networks is discussed in detail in Section III. Section IV discusses about proposed three-voltage supply systems and Section V deals with Anti-resonance effects of decoupling capacitors. Some specific conclusions are summarized in Section IV.

## II. POWER DISTRIBUTION NETWORKS

The Power distribution network in a chip consists of chip level power distribution with thin-oxide decoupling capacitors, the package level power distribution with planes and mid frequency decoupling capacitors and board level power distribution with planes, low-frequency decoupling capacitors and voltage regulator module. With all these parameters involved in drawing, the design of power distribution system has become an increasingly difficult challenge in modern CMOS circuits [5].

Power distribution system plays a major role in determining the power consumption of the whole circuit. In a design of most microprocessors or ASIC chips, the target market sets the operating frequency. The timing constraints in the chip are in turn set by the operating frequency. Designers need to optimize the design to reduce the power consumption with the specified timing constraints. If the supply voltage is reduced under the constant $V_{th}$, the critical path delay will not meet the timing constraints [6]. As CMOS technologies are scaled, the power supply voltage is lowered. With the general scaling theory, the current I is increasing and the power supply voltage is decreasing. The impedance of the power distribution system should, therefore, be decreased to satisfy to satisfy power noise constraints. The target impedance of a power distribution system is falling at an alarming rate, a factor of five per computer generation.

### III. PRIOR WORK

Recently, a model of the impedance of a power distribution system with two supply voltages was studied [1,3]. Impedance model of the power distribution system with two supply voltages is shown in the Fig. 1. The impedance of the network can be calculated as

$$Z = \frac{Z_1 Z_{12} + Z_1 Z_2}{Z_1 + Z_{12} + Z_2} \quad (1)$$

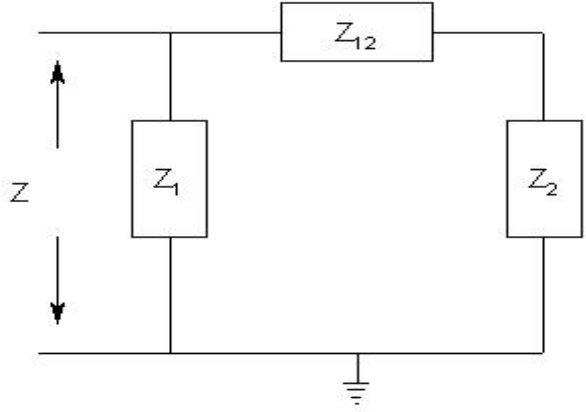

Fig 1. Impedance of power distribution system with two supply voltages seen from the load of the power supply $V_{d1}$

Decoupling capacitors have traditionally been modeled as a series resistance-inductance-capacitance (RLC) network. Fig.2 represents the schematic representation of the power distribution network with two voltage supplies and the decoupling capacitors represented by RLC series networks. Upon calculations the impedance of the network turns out to be

$$Z = \frac{a_4 s^4 + a_3 s^3 + a_2 s^2 + a_1 s + a_0}{b_3 s^3 + b_2 s^2 + b_1 s} \quad (2)$$

where

$a_4 = L_1(L_{12} + L_2)$
$a_3 = R_1 L_{12} + R_{12} L_1 + R_1 L_2 + R_2 L_1$
$a_2 = R_1 R_{12} + R_1 R_2 + (L_1/C_{12}) + (L_{12}/C_1) + (L_1/C_2) + (L_2 C_1)$
$a_1 = (R_1/C_2) + (R_2/C_1) + (R_1/C_{12}) + (R_{12}/C_1)$
$a_0 = (C_{12} + C_2)/C_1 C_2 C_3$
$b_3 = L_1 + L_{12} + L_2$
$b_2 = R_1 + R_{12} + R_2$
$b_1 = (1/C_1) + (1/C_{12}) + (1/C_2)$

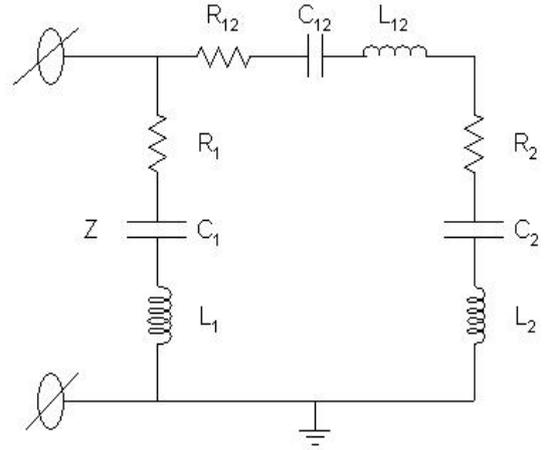

Fig.2. Impedance of power distribution system with two supply voltages and the decoupling capacitors represented as series RLC networks.

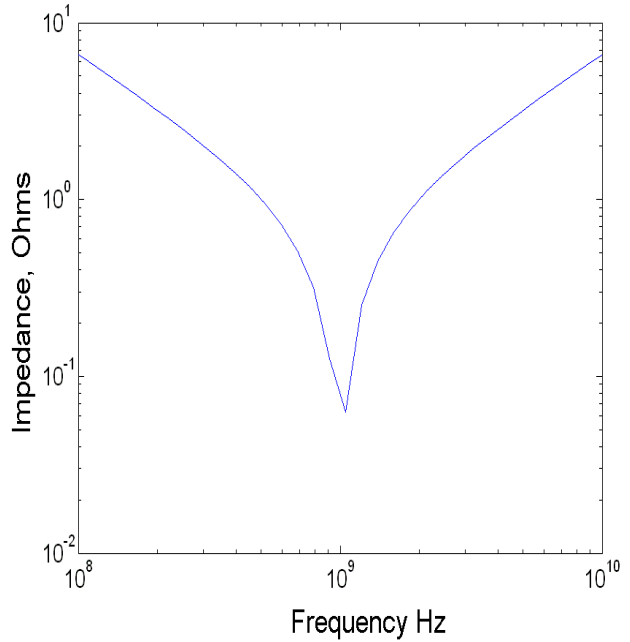

Fig. 3. Frequency dependence of the impedance of a power distribution system with dual supply voltages, $R_1=R_{12}=R_2 = 10m\Omega$, $C_1=C_{12}=C_2=1nF$ and $L_1=L_{12}=L_2=1nH$. All the parameters of the power distribution network are set to identical values; the system behaves as a single capacitor with a minimum at the resonant frequency. The minimum power distribution system impedance is limited by the ESR of the decoupling capacitors.

Fig.3.illustrates the frequency dependence of the impedance of the power distribution system with dual voltage supplies. Here the results obtained are contradictory with the results reported by Popovich et. al [1]. For the same values of the components (R, L and C), the resonance peak lies at 1GHz where as Popovich et. al obtained the resonance peak at 158.5MHz. However, the nature of the curve is the approximate the same even in [1].

## IV. THREE VOLTAGE SUPPLY SYSTEM

Unprecedented increase in the growth of semiconductor industry has given rise to innovative designs such as SoC integration and NoC Systems. The Complexity of System-on-Chip (SoC) has ever increased due to incorporation of both digital and analog modules. Typically energy requirements of the modules vary with the functionality. The energy requirements of digital cores might vary with the analog cores. To accomplish such energy requirements and to maintain the high performance, techniques such as multiple power supplies are proposed. Certain high-performance applications do require three voltage supply systems, one for critical paths, second for lesser critical paths and third supply for least critical paths. The complexity of the power distribution system for three supply voltages is much higher compared to two supply voltages. Fig. 4 shows the three-voltage supply schematic. Each voltage supply is decoupled with the other voltage supply using decoupling capacitors.

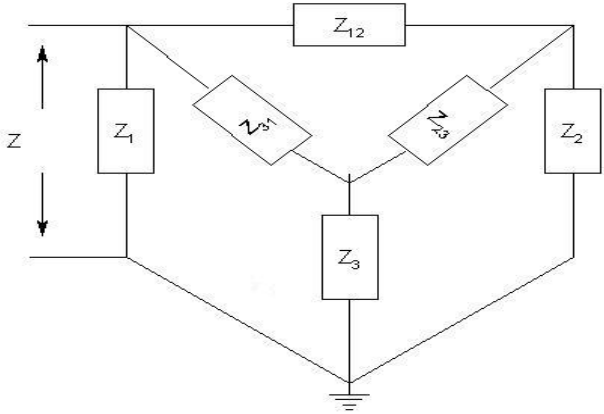

Fig. 4. :Impedance of power distribution system with three supply voltages seen from the load of the power supply $V_{d1}$

Using RLC series network to represent decoupling capacitors, the power distribution systems transforms into Fig. 5. Using Delta to Wye Transformations the impedance of the circuit is evaluated. The final equation obtained is given by (3). To analyze the frequency dependence of the impedance of the power distribution system with three voltage supplies, there is a necessity to make certain approximations. These approximations are made to simply the tedious calculation involved in solving (3). The approximating used here is that all the decoupling impedances $Z_{12}$, $Z_{23}$ and $Z_{31}$ are identical and equal to $Z_0$, where $Z_0 = R_0 + sL_0 + (1/sC_0)$. Substituting this approximation, (3) transforms into (4).

The frequency dependence of the closed form expression for the impedance of a power distribution system with dual power supply voltages is illustrated in the Fig. 6. Effective series impedance limits the power efficiency of the network. In context of on-chip applications, ESR includes the parasitic resistance of the decoupling capacitor and the resistance of the power distribution network connecting a decoupling capacitor to a load. Assuming that the supply voltage is around 1.25V and drives a current of 80A in 90nm and allowed ripple is about 5%, the obtained the target impedance of the power distribution system is

$$Z_{target} = \frac{Vdd \times r}{I} = \frac{1.25 volts \times 5\%}{80 amperes} \approx 0.00078 ohms$$

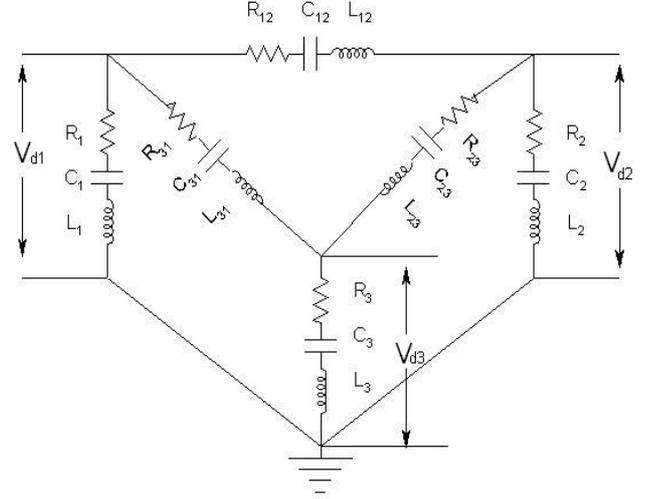

Fig. 5. Impedance of power distribution system with three supply voltages and the decoupling capacitors represented as series RLC networks.

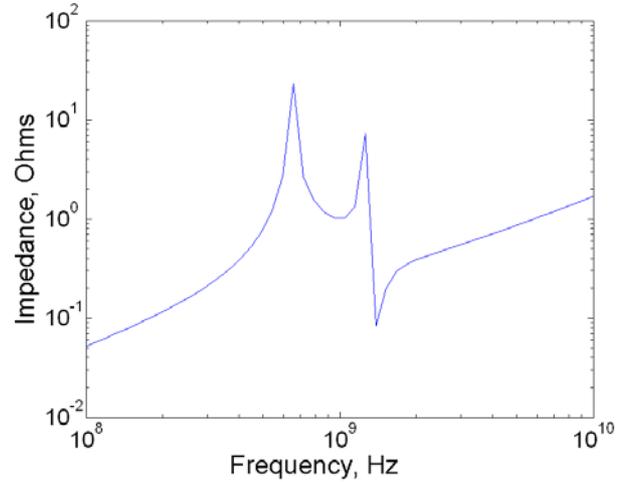

Fig. 6. Frequency dependence of the impedance of a power distribution system with dual supply voltages, R1 = R2 =R3 =R0 = 10mΩ, C1 = C2 = C3 = C0 = 1nF and L1 = L2 = L3= L0 = 1nH.

Multiple decoupling capacitors are placed in the power distribution hierarchy to maintain the $Z_{target}$ as described above. To maintain the impedance of the power distribution system below a specified level it is necessary to make sure that any variation in the frequency doesn't affect the impedance. At least the drastic variations in the impedance with frequency should be removed in the range of consideration.

$$Z = \frac{Z_1Z_2Z_3Z_{IJ}^2 + Z_1Z_{IJ}Z_{23}(Z_2Z_{31}+Z_3Z_{12})+Z_1Z_{23}^2Z_{12}Z_{13}+Z_1Z_{12}Z_{31}[(Z_2+Z_3)Z_{IJ}+Z_{23}Z_{12}+Z_{31}Z_{23})]}{Z_2Z_3Z_{IJ}^2 + Z_{IJ}Z_{23}(Z_2Z_{31}+Z_3Z_{12})+Z_{23}^2Z_{12}Z_{31}+(Z_1Z_{IJ}+Z_{12}Z_{31})[(Z_2+Z_3)Z_{IJ}+Z_{23}Z_{12}+Z_{31}Z_{23})]} \quad (3)$$

$$Z = \frac{a_6s^6+a_5s^5+a_4s^4+a_3s^3+a_2s^2+a_1s+a_0}{b_5s^5+b_4s^4+b_3s^3+b_2s^2+b_1s} \quad (4)$$

Where,
$a_6 = L_1L_0(L_2^1+L_3^1)$
$a_5 = L_2^1L_3^1 + R_1L_0(L_2^1+L_3^1) + L_1R_0(L_2^1+L_3^1) + L_1L_0(R_2^1+R_3^1)$
$a_4 = R_2^1L_3^1 + L_2^1L_3^1 + R_0R_1(L_2^1+L_3^1) + R_1L_0(R_2^1+R_3^1) + L_1R_0(R_2^1+R_3^1) + L_1L_0(1/C_3^1+1/C_2^1)$
$\quad + L_1(L_2^1+L_3^1)/C_0 + L_0(L_2^1+L_3^1)/C_0$
$a_3 = R_2^1R_3^1 + L_2^1/C_3^1 + L_3^1/C_2^1 + R_0R_1(R_2^1+R_3^1) + R_1L_0(1/C_3^1+1/C_2^1) + R_1/C_0(L_2^1+L_3^1) +$
$\quad L_1R_0(1/C_3^1+1/C_2^1)+L_1/C_0(R_2^1+R_3^1)+R_0/C_1(L_2^1+L_3^1)+L_0/C_1(R_3^1+R_2^1)$
$a_2 = R_2^1/C_3^1 + R_3^1/C_2^1 + R_0R_1(1/C_3^1+1/C_2^1) + R_1(R_2^1+R_3^1)/C_0 + L_1/C_0(1/C_3^1+1/C_2^1)$
$\quad + L_0/C_1(1/C_3^1+1/C_2^1) + R_0/C_1(R_2^1+R_3^1) + (L_2^1+L_3^1)/C_1C_0$
$a_1 = 1/C_3^1C_2^1 + R_1/C_0(1/C_3^1+1/C_2^1) + R_0/C_1(1/C_3^1+1/C_2^1) + (R_2^1+R_3^1)/C_1C_0$
$a_0 = 1/C_0C_1(1/C_3^1+1/C_2^1)$
$b_5 = L_2^1L_3^1 + L_1^1(L_2^1+L_3^1)$
$b_4 = R_2^1L_3^1 + L_2^1R_3^1 + R_1^1(L_2^1+L_3^1) + L_1^1(R_2^1+R_3^1)$
$b_3 = (L_2^1+L_3^1)/C_1^1 + R_2^1R_3^1 + L_2^1/C_3^1 + L_3^1/C_2^1 + R_1^1(R_2^1+R_3^1) + L_1^1(1/C_3^1+1/C_2^1)$
$b_2 = R_1^1(1/C_3^1+1/C_2^1) + (R_2^1+R_3^1)/C_1^1 + R_2^1/C_3^1 + R_3^1/C_2^1$
$b_1 = 1/C_1^1(1/C_3^1+1/C_2^1) + 1/C_3^1C_2^1$

$R_1^1=3R_1+R_0;$    $R_2^1=3R_2+R_0;$    $R_3^1=3R_3+R_0$
$L_1^1=3L_1+L_0;$    $L_2^1=3L_2+L_0;$    $L_31=3L_3+L_0$
$1/C_1^1=(3/C_1)+1/C_0;$    $1/C_2^1=(3/C_2)+1/C_0;$    $1/C_3^1=(3/C_3)+1/C_0$

## V. ANTI RESONANCE OF PARALLEL CAPACITORS

In two voltage supplies power distribution hierarchy, the values of the components considered in Fig. 3. were equal. Having equal values of the components cancels out the poles and zeros of the equation (2), resulting in pure resonance as seen in Fig. 3. In the Fig. 3 two zeros are located at the same frequency as the pole. The pole is, therefore, cancelled for this special case and the circuit behaves as a series RLC circuit with one resonant frequency. However, in practical circuits the values of components are seldom equal. In that case poles are not cancelled with zeros. For instance, the value of $C_{12}$ is not equal to $C_1$ or $C_2$, the smooth resonance is disturbed due to the existence of additional poles and zeros. With two poles, two different resonant frequencies are obtained, $f_1$ and $f_2$, and due to existence of one additional zero a peak is observed between $f_1$ and $f_2$. Occurrence of such peaks is called as anti-resonance effect of parallel capacitors. Fig. 7 shows the anti-resonance plot for two voltage supplies power distribution.

However, the scenario is not the same in three level voltage supply hierarchy. Equal values of all the components do not cancel out two poles so that only one resonant peak can occur. This can be observed in Fig. 6 where there are two anti-resonance peaks and two resonant peaks. To obtain single resonance requires calculation of component values that cancel out all the poles leaving only one pole.

Anti-resonance is highly undesirable phenomenon because it causes the unexpected rise in the impedance of the power distribution network. To cancel the anti-resonance effect at a given frequency, a small decoupling capacitor is place in parallel, shifting the anti-resonance spike to a higher frequency. This procedure is repeated until the anti-resonance spike appears at a frequency out of range of the operating frequencies of the system. Another technique for shifting the anti-resonance spike to a higher frequency is to decrease the effective series inductance (ESL) of the decoupling capacitor. To shift the poles to a higher frequency, the ESL of the decoupling capacitors must be decreased. If the ESL of the decoupling capacitors is close to zero, the impedance of a power delivery network will not produce overshoots over a wide range of operating frequencies. The dependence of the power distribution system impedance on the ESL is shown in Fig. 8. By lowering the system inductance, the quality factor is decreased. The peaks become wider in frequency and lower in magnitude. The amplitude of the anti-resonant spikes can be decreased by lowering the ESL of all of the decoupling capacitors within the power distribution system.

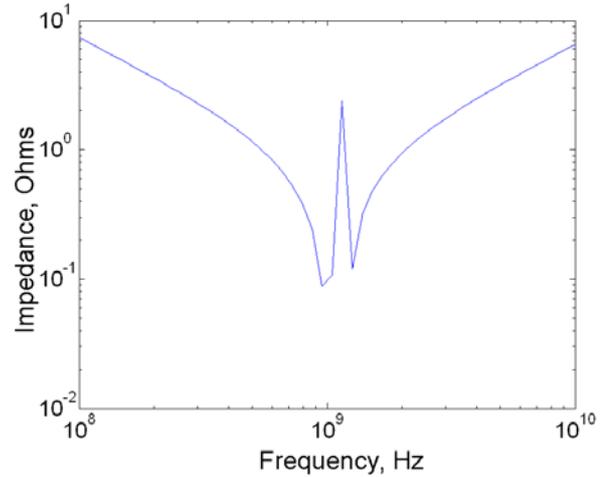

Fig. 7. Frequency dependence of the impedance of a power distribution system with dual supply voltages, $R_1=R_{12}=R_2 = 10m\Omega$, $C_1=C_2=1nF$, $C_{12} = 0.5nF$ and $L_1=L_{12}=L_2=1nH$. The peak in the middle shows the occurrence of anti-resonance.

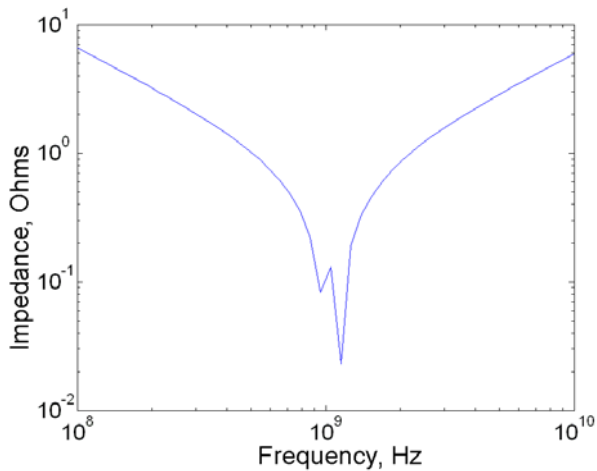

Fig. 8. Anti-resonance of a power distribution system with dual voltage supply voltages. $R_1=R_{12}=R_2 = 10m\Omega$, $C_1=C_{12}=C_2=1nF$ and $L_1=L_2=1nH$ and $L_{12} = 0.5nH$. Reduction in $L_{12}$ has resulted in decrease in the amplitude of the anti-resonant frequency.

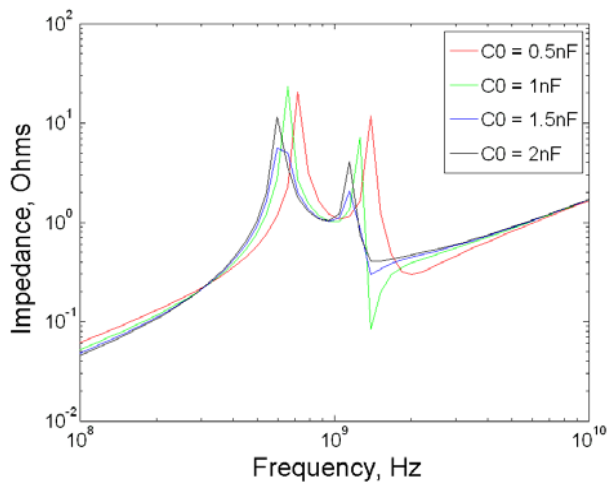

Fig.9. Shift in the peaks of anti-resonance towards left by increasing the decoupling capacitance. The decoupling capacitance $C_0$ used is varied from 0.5nF to 2nF.

## VI. CONCLUSION

It has become common practice to use multiple on-chip power supply voltages to reduce power distribution without degrading system speed. To maintain the impedance of a power distribution system below specified impedance, multiple decoupling capacitors are placed at different levels of the power grid hierarchy. The decoupling capacitors are placed both with progressively decreasing values to shift the anti-resonance spike beyond the maximum operating frequency, and with increasing ESR to control the damping characteristics. Increasing ESR also broadens the frequency range of the anti-resonant spikes, ensuring that the amplitude of the output impedance remains below the target impedance. Another strategy is to limit the magnitude of the anti-resonant spikes by reducing the ESL of all the decoupling capacitors. Also the equations for three level power distribution hierarchies were calculated and simulated.